\documentclass[a4paper, aps, amsmath, 10pt, nofootinbib]{revtex4-2}
\usepackage[left=3cm,right=3cm,top=3cm,bottom=3cm]{geometry}

\usepackage{graphicx}
\usepackage{booktabs}
\usepackage{tikz}
\usetikzlibrary{arrows.meta}
\usepackage{pgfplots}
\pgfplotsset{compat=1.14}
\usepackage{anyfontsize}
\usepackage{xcolor}
\usepackage{siunitx}
\usepackage{physics}
\usepackage{amsfonts}
\usepackage{mathtools}
\usepackage{svg}
\usepackage{hyperref}
\usepackage[skip=2pt,font=normalsize]{subcaption}

\begin{document}

\title{The Casimir energy with perfect electromagnetic boundary conditions and duality: a field-theoretic approach}
\author{David Dudal}
\email{david.dudal@kuleuven.be}
\affiliation{KU Leuven Campus Kortrijk -- Kulak, Department of Physics, Etienne Sabbelaan 53 bus 7657, 8500 Kortrijk, Belgium}

\author{Aaron Gobeyn}
\email{aaron.gobeyn@student.kuleuven.be}
\affiliation{KU Leuven Campus Kortrijk -- Kulak, Department of Physics, Etienne Sabbelaan 53 bus 7657, 8500 Kortrijk, Belgium}

\author{Thomas Oosthuyse}
\email{thomas.oosthuyse@kuleuven.be}
\affiliation{KU Leuven Campus Kortrijk -- Kulak, Department of Physics, Etienne Sabbelaan 53 bus 7657, 8500 Kortrijk, Belgium}

\author{Sebbe Stouten}
\email{sebbe.stouten@kuleuven.be}
\thanks{corresponding author}
\affiliation{KU Leuven Campus Kortrijk -- Kulak, Department of Physics, Etienne Sabbelaan 53 bus 7657, 8500 Kortrijk, Belgium}

\author{David Vercauteren}
\email{vercauterendavid@duytan.edu.vn}
\affiliation{Institute of Research and Development, Duy Tan University, Da Nang 550000, Vietnam}
\affiliation{Faculty of Natural Sciences, Duy Tan University, Da Nang 550000, Vietnam}

\begin{abstract}
	Using functional integral methods, we study the Casimir effect for the case of two infinite parallel plates in the QED vacuum, with (different) perfect electromagnetic boundary conditions applied to both plates. To enforce these boundary conditions, we add two Lagrange multiplier fields to the action. We subsequently recover the known Casimir energy in two ways: once directly from the path integral, and once as the vacuum expectation value of the $00$-component of the energy-momentum tensor. Comparing both methods, we show that the energy-momentum tensor must be modified, and that it picks up boundary contributions as a consequence. We also discuss electromagnetic duality-invariance of the theory and its interplay with the boundaries by generalizing the Deser--Teitelboim implementation of the duality transformation.
\end{abstract}

\maketitle

\section{Motivation}
The Casimir effect is a well-known macroscopic manifestation of microscopic quantum vacuum fluctuations interacting with a spatiotemporal boundary \cite{Plunien:1986ca,Bordag:2001qi,Milton:2004ya,Bordag:2009zz}, that has been verified experimentally (e.g.~\cite{Deriagin:1956zt,Black1960,Lamoreaux1997,Mohideen:1998iz,Bressi:2002fr}). Intuitively, one can expect the boundary to restrict the possible field configurations, such that the amount of allowed vacuum states can vary by region. Let us consider QED in 4D with the simplest non-trivial boundary: two parallel, conductive plates. In this case, we have fewer vacuum states inside the plates than outside, resulting in a net force pushing them together \cite{Casimir:1948dh}. 

However, some caution is needed regarding this naive picture, as parallel plate configurations exist for which the net Casimir force is repulsive instead of attractive \cite{Boyer:1974}, even at finite temperature \cite{daSilva:2001yp}. The no-go theorem \cite{Kenneth:2006vr} tells us that such configurations must either break reflection symmetry (e.g.\ using reflection-breaking media such as chiral materials \cite{Jiang:2018ivv,Fukushima:2019sjn,Canfora:2022xcx,Nakayama:2023zvm,Ema:2023kvw,Favitta:2023hlx}), or must use plates made of more exotic materials than dielectrics and perfect electric conductors (PEC). Interesting examples of such exotic materials are topological insulators \cite{Qi:2008ew,Grushin:2010qoi,Tajik:2022kka}, Weyl semimetals \cite{Grushin:2012mt,Oosthuyse:2023mbs} and magnetically permeable materials \cite{Boyer:1974}. The latter might be realized experimentally using meta-materials \cite{Lopez:2022bky}, and they can be modeled mathematically by perfect magnetic conductors (PMC) \cite{Edery:2008rj}. The boundary conditions that we will consider here, are a linear combination of PECs and PMCs: perfect electromagnetic conductors (PEMC) \cite{lindell2005perfect,Rode:2017yqy,Schoger:2024zig}, which combine conducting and permeable properties.

In this paper, we will discuss the Casimir effect for PEMC plates using a functional integral formalism. We reproduce the known one-loop expression for the Casimir energy, which has been calculated in \cite{Rode:2017yqy} using a scattering matrix approach. In fact, we will present two slightly different methods for this calculation. The first one has been developed in \cite{Dudal:2020yah,Canfora:2022xcx,Oosthuyse:2023mbs}, and only requires the path integral. This method enforces the boundary conditions in an explicitly gauge invariant fashion and yields a 3D non-local effective boundary theory, meaning that the Casimir effect can be interpreted as a type of boundary dynamics \cite{Juarez-Aubry:2020psk}. The second one additionally makes use of the energy-momentum tensor (EMT), as was done in \cite{Bordag:1983zk} for PEC plates. Following this method, we find interesting new boundary contributions to the EMT that have to be taken into account. Those had not been noticed before, because their contribution is zero in the PEC case studied in \cite{Bordag:1983zk}, where the standard Maxwell EMT was employed.

The PEMC boundaries also naturally lead one to consider electromagnetic duality-invariance of the theory \cite{deser1976duality} and its interplay with the boundary \cite{lindell2005perfect}. We have reformulated this duality transformation on the level of the gauge field by suitably adapting the formalism of \cite{deser1976duality} in the presence of these PEMC plates.

The present paper is organized as follows. In section \ref{ch:setup}, the considered setup is specified and duality-invariance of the system is discussed. After this, we derive the 3D non-local effective boundary action in section \ref{ch:3Dtheory} and describe both functional integral methods in section \ref{ch:methods}. Finally, some concluding remarks are presented in section \ref{ch:conclusion}.

\section{Setup}\label{ch:setup}
\subsection{Action and boundary conditions}
Since we wish to calculate the leading term of the Casimir energy in QED, it suffices to consider the gauge field: the fermion field only contributes sub-leading terms \cite{Bordag:1983zk}. For notational ease, we will work with the Wick-rotated Euclidean action from the start, which is always possible for non-dynamical systems. Let us thus consider the Euclidean Maxwell action 
\begin{equation}\label{eq:MaxwellAction}
	S_\text{Maxwell} = \int \dd[4]{x} \frac{1}{4} F_{\mu\nu} F_{\mu\nu},
\end{equation}
in which \( F_{\mu\nu} = \partial_\mu A_\nu - \partial_\nu A_\mu\) is the Maxwell field strength tensor. Since we work in Euclidean space, we will simply write all spacetime indices as lower indices, and understand repeated indices to be summed over. In order to fix the gauge invariance under \(A_\mu \rightarrow A_\mu + \partial_\mu \chi\), we add a linear gauge fixing term to the action
\begin{equation*}
	S_\text{GF} = \int \dd[4]{x} \left( i h \partial_\mu A_\mu + \frac{\xi}{2} h^2 \right)
\end{equation*}
where \(\xi\) is the gauge parameter, and \(h\) the Nakanishi-Lautrup multiplier field \cite{Nakanishi:1966zz,Lautrup1967canonical,Nakanishi:1990qm}.
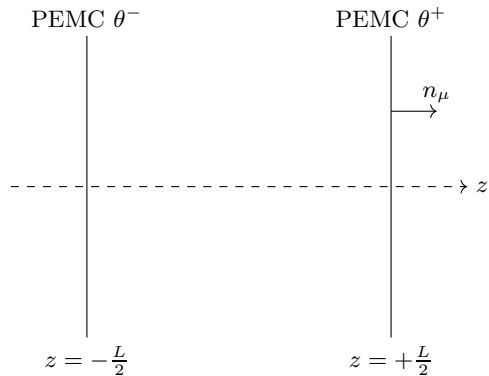
\begin{figure} 
	\centering
	\begin{tikzpicture}[scale=1]
		\draw[dashed,->] (0,0) -- (6,0) node[right] {$z$};
		\draw (1,2) node[above] {PEMC \(\theta^-\)} -- (1,-2) node[below] {$z=-\frac{L}{2}$};
		\draw (5,2) node[above] {PEMC \(\theta^+\)} -- (5,-2) node[below] {$z=+\frac{L}{2}$};
		\draw[->] (5,1) -- (5.6,1) node[above] {\(n_\mu\)};
	\end{tikzpicture}
	\caption{Schematic representation of the boundary configuration: infinitely large plates at \(z=\pm \frac{L}{2}\), with PEMC boundary conditions applied with duality angle \(\theta^\pm\).  }
	\label{fig:boundary}
\end{figure}

The geometry we will discuss is the following: 4-dimensional Euclidean space with two infinitely large, infinitely thin, parallel plates placed at \(z=-\frac{L}{2}\) and \(z=+\frac{L}{2}\) respectively (\(L>0\)) (see FIG. \ref{fig:boundary}). Let us also introduce the normal vector \(n_\mu = (0,0,0,1)\). In order to define our boundary conditions, we introduce the dual field strength tensor \(\widetilde{F}\) via 
\begin{equation*}
	\widetilde{F}_{\mu\nu} = \frac{i}{2} \varepsilon_{\mu\nu\alpha\beta} F_{\alpha\beta},
\end{equation*}
in which a factor \(i\) has appeared because of Wick-rotating the Levi-Civita tensor.

Now we can define the boundary conditions\footnote{Technically speaking, the boundary conditions imposed at the plates do not correspond to conditions at the ``end of space''; rather the plates are interfaces placed in a higher-dimensional bulk. The more general study of quantum field theories in presence of defects has seen an increased interest during recent years, mostly in the context of conformal field theories and/or holography \cite{Andrei:2018die}. } corresponding to perfect electric conductors (PEC) and perfect magnetic conductors (PMC), respectively defined by (\ref{eq:PEC}) and (\ref{eq:PMC}). 
\begin{align}
	\text{PEC}: \quad &\widetilde{F}_{\mu\nu}n_\nu = 0 \label{eq:PEC}\\
	\text{PMC}: \quad &F_{\mu\nu}n_\nu = 0 \label{eq:PMC}
\end{align}
In this paper, we will however consider slightly more general boundary conditions, namely perfect electromagnetic conductors (PEMC), being simply a linear combination of PEC and PMC. We will parameterize this linear combination using a duality angle \(\theta\) (the name of which is justified in Section \ref{ch:duality}):
\begin{equation*}
	\text{PEMC}: G_\mu(\theta) \coloneq \left(\cos (\theta) F_{\mu \nu} + \sin (\theta) \widetilde{F}_{\mu \nu}\right) n_\nu = 0,
\end{equation*}
such that we retrieve the PEC (resp.\ PMC) conditions for \(\theta = \frac\pi2\) (resp.\ \(\theta=0\)). PEMC conditions are the most general set of linear boundary conditions exhibiting gauge invariance \cite{Vassilevich:2003xt}. As indicated in FIG. \ref{fig:boundary}, we will apply PEMC conditions with duality angle \(\theta^\pm\) on \(z=z^\pm \coloneq \pm\frac{L}{2}\):
\begin{equation}\label{eq:bndCond}
	\left. G_\mu(\theta^-) \right|_{z=z^-} = 0, \qquad \left. G_\mu(\theta^+) \right|_{z=z^+} = 0.
\end{equation}

Since we wish to work in a field-theoretic formalism, we now need to incorporate the boundary conditions (\ref{eq:bndCond}) into the action. Following the explicitly gauge-invariant methodology developed in \cite{Dudal:2020yah,Canfora:2022xcx,Oosthuyse:2023mbs}, we do this by introducing a pair of Lagrange multiplier fields \(b^\pm\). If we define \(\Sigma^\pm\) as the 3-dimensional subspace described by \(z = z^\pm\), the boundary term in the action becomes
\begin{equation*}
	S_\text{BC} = \int_{\Sigma^-}\dd[3]{\mathbf{x}} b^-_i(\mathbf{x}) G_i(\theta^-) + \int_{\Sigma^+}\dd[3]{\mathbf{x}} b^+_i(\mathbf{x}) G_i(\theta^+),
\end{equation*} 
where the \(b^\pm\)-fields only carry a Latin index \(\in \{t,x,y\}\) (because \(G_z \equiv 0\) automatically) and only depend on the 3-vector \(\mathbf{x} = (t,x,y)\).
We will introduce some extra notation in order to get more compact expressions in what follows. Let us introduce an anti-symmetrized boundary tensor
\begin{equation*}
	b_{\mu\nu}^\gamma \coloneq \left( b_\mu^\gamma n_\nu -b_\nu^\gamma n_\mu \right),
\end{equation*}
where we have introduced a plate index \(\gamma\in\{-,+\}\), and in which we understand that \(b_z^\gamma \coloneq 0\). 
Its corresponding dual is given by 
\begin{equation*}
	\tilde{b}_{\mu\nu}^\gamma = \frac{i}{2} \varepsilon_{\mu\nu\alpha\beta} b_{\alpha\beta}^\gamma.
\end{equation*}
We can then combine both plates in a single anti-symmetric tensor
\begin{equation}\label{eq:defH}
	H_{\mu\nu} \coloneq  \delta(z-z^\gamma) \left( b_{\mu\nu}^\gamma \cos(\theta^\gamma) + \tilde{b}_{\mu\nu}^\gamma \sin(\theta^\gamma) \right),
\end{equation}
where we understand repeated appearances of plate indices to be summed over.\footnote{In contrast to spacetime indices, the plate indices need not appear in pairs, but can also form triplets, quadruplets, etc.} Using this tensor, the boundary action simply becomes
\begin{equation*}
	S_\text{BC} = \int \dd[4]{x} \frac12 H_{\mu\nu} F_{\mu\nu}.
\end{equation*}
The (Euclidean) action we will be studying in this paper is thus given by
\begin{align}\label{eq:action}
	\begin{split}
	S &= S_\text{Maxwell} + S_\text{GF} + S_\text{BC} \\
	&=  \int \dd[4]{x} \left[ \frac{1}{4} F_{\mu\nu} F_{\mu\nu} + \left( i h \partial_\mu A_\mu + \frac{\xi}{2} h^2 \right) + \frac12 H_{\mu\nu} F_{\mu\nu}\right].
	\end{split}
\end{align}

\subsection{Electromagnetic duality-invariance}\label{ch:duality}
Electromagnetic duality transformations are defined as rotations of the electric and magnetic field into each other, or, equivalently, as rotations of the Maxwell field tensor and its dual into each other.  It is well known that the Maxwell equation of motion is invariant under such transformations, although the action (\ref{eq:MaxwellAction}) changes by a scale factor. As is discussed in \cite{deser1976duality}, for duality transformations to be a meaningful symmetry, it is crucial that they can be implemented at the level of the fundamental field(s). 

In Euclidean space without boundaries, an infinitesimal duality transformation is given by 
\begin{equation*}
	\begin{cases}
		\delta \mathbf{E} = i\dd\alpha \mathbf{B} \\
		\delta \mathbf{B} = i\dd\alpha \mathbf{E}
	\end{cases},
\end{equation*}
or, equivalently, by 
\begin{equation}\label{eq:dualRotF}
	\begin{cases}
		\delta F_{\mu\nu} = -i\dd\alpha\widetilde{F}_{\mu\nu} \\
		\delta \widetilde{F}_{\mu\nu} = -i\dd\alpha F_{\mu\nu}
	\end{cases}.
\end{equation}
Generalizing \cite{deser1976duality}, and using the equations of motion for the photon field as obtained from the \emph{full action} (\ref{eq:action})
\begin{equation*}
	\begin{cases}
		(\partial_0^2+\nabla^2)A_\nu= i\partial_\nu h + \partial_\nu H_{\mu\nu}  \\
		i \partial_\mu A_\mu = -\xi h
	\end{cases},
\end{equation*}
the interested reader can verify that, irrespective of \(\xi\), the transformation of the fundamental field
\begin{equation}
	\begin{cases}
		\delta A_0 &= 0 \\
		\delta A_i &= i\dd\alpha \frac{1}{\nabla^2}\varepsilon_{ijk}\partial_j\partial_0 A_k 
	\end{cases}
\end{equation}
implements a transformation of the field strength similar to (\ref{eq:dualRotF}), but containing a boundary contribution:
\begin{equation}\label{eq:dualRotFBoundary}
	\begin{cases}
		\delta F_{\mu\nu} = -i\dd\alpha \left( \widetilde{F}_{\mu\nu} - \delta(z-z^\gamma)X_{\mu\nu}^\gamma \right)\\
		\delta \widetilde{F}_{\mu\nu} = -i\dd\alpha \left( F_{\mu\nu} - \delta(z-z^\gamma)\widetilde{X}_{\mu\nu}^\gamma \right)
	\end{cases},
\end{equation}
where \(X_{\mu\nu}^\gamma\) only depends on the boundary fields \(b_i^\gamma\) (non-locally). On the boundary fields and the Nakanishi-Lautrup field, the trivial transformations \(\delta b_i^\gamma=0\) and \(\delta h = 0\) are imposed. 

We now show term by term that the full action \(S = S_\text{Maxwell} + S_\text{GF} + S_\text{BC}\) is invariant under this symmetry (\ref{eq:dualRotFBoundary}). After partial integration, \(S_\text{Maxwell}\) is proportional to \(\int A_\mu \partial^2 A_\mu\), the variation of which is proportional to \(\int A_i \frac{\partial^2}{\nabla^2}\varepsilon_{ijk} \partial_j \partial_0 A_k\). This expression equals minus itself after partial integration of \(\partial_0\) and \(\partial_j\), and thus equals zero. The variation of \(S_\text{GF}\) is immediately seen to be zero because \(\varepsilon\) is anti-symmetric. This leaves us with \(S_\text{BC}\). Setting \(\delta \theta^\gamma = \dd\alpha\) \cite{lindell2005perfect} and using (\ref{eq:dualRotFBoundary}), we find that \(\delta G_\mu(\theta) = \delta \left(\cos (\theta) F_{\mu \nu} + \sin (\theta) \widetilde{F}_{\mu \nu}\right) n_\nu \propto \delta(z-z^\lambda)\). If we now vary \(S_\text{BC} = \int \dd[4]{x} \delta(z-z^\gamma) b_i^\gamma G_i(\theta^\gamma)\), we find that it contains \(\delta(z-z^\gamma) \delta(z-z^\lambda)\), which yields zero if \(\gamma \neq \lambda\), and \(\delta(0)\) if \(\gamma=\lambda\), such that it also vanishes in dimensional regularization. (Alternatively, one can regularize by adding a counterterm to the action.)

We have thus carefully checked that duality transformations are a symmetry of the system including boundaries. Therefore, all physical observables must be duality-invariant. In particular, the Casimir energy is to be invariant under adding a fixed angle to both PEMC angles \(\theta^\pm\), meaning it may only depend on \(\delta \coloneq \theta^+ - \theta^-\). Moreover, recovering this duality-invariant result has led us to construct the corrected EMT, see Section \ref{ch:EMT}.

\section{Reduction to a 3D non-local effective boundary theory}\label{ch:3Dtheory}
Our main object of study will be the functional integral of the total action (\ref{eq:action})
\begin{equation}\label{eq:pathInt}
	Z = \int \mathcal{D}\!A\, \mathcal{D}b^-\, \mathcal{D}b^+\, e^{-S}.
\end{equation}
When evaluating this integral, one has the choice to integrate out the photon field first (as in \cite{Golestanian_1998,Dudal:2020yah}), or the boundary fields first (as e.g.\ in \cite{Bordag:1983zk}). We will go ahead by integrating out the photon field, since this results in a simpler computation for the Casimir energy later on. This will leave us with a 3D non-local effective boundary theory, which we will derive in this section.

Going on-shell for the \(h\)-field and using partial integration to write the action (\ref{eq:action}) as a quadratic form, we get 
\begin{equation*}
	S = -\frac12 \int \dd[4]{x} \left[ A_\mu \left( \delta_{\mu\nu} \square - \big(1 - \frac{1}{\xi}\big) \partial_\mu\partial_\nu\right) A_\nu -2 \left( \partial_\mu H_{\mu\nu} \right) A_\nu \right].
\end{equation*}
Because of the translational symmetry of the action, going to Fourier space will make computations easier (see Appendix \ref{ch:appendix} for our conventions). In Fourier space, our action takes the form 
\begin{equation}\label{eq:FourierAction}
	S = \frac12 \int \frac{\dd[4]{k}}{(2\pi)^4} \bigg[ \hat{A}_\mu(k) K_{\mu\nu}^{(\xi)} \hat{A}_\nu(-k) - 2 i k_\mu \hat{H}_{\mu\nu}(k) \hat{A}_\nu(-k) \bigg].
\end{equation}
The quadratic form in the first term is 
\begin{equation*}
	K_{\mu\nu}^{(\xi)} = \left( k^2\delta_{\mu\nu}  - \big(1 - \frac{1}{\xi}\big) k_\mu k_\nu\right),
\end{equation*}
where we explicitly denote its \(\xi\)-dependence in the superscript. The Fourier transform of \(H_{\mu\nu}\) in the second term is given by
\begin{equation}\label{eq:defHFourier}
	\hat{H}_{\mu\nu}(k) = e^{i k_z z^\gamma} \left( \hat{b}_{\mu\nu}^\gamma(\mathbf{k}) \cos(\theta^\gamma) + \hat{\tilde{b}}_{\mu\nu}^\gamma(\mathbf{k}) \sin(\theta^\gamma) \right).
\end{equation}
This action (\ref{eq:FourierAction}) is clearly of the form
\begin{equation}\label{eq:quadraticForm}
	S = \frac12 \int \frac{\dd[4]{k}}{(2\pi)^4} \bigg[ \hat{A}_\mu(k) K_{\mu\nu}^{(\xi)} \hat{A}_\nu(-k) - \hat{v}_\nu(k) \hat{A}_\nu(-k) \bigg],
\end{equation}
with the vector \(\hat{v}_\nu(k) \coloneq 2 i k_\mu \hat{H}_{\mu\nu}(k) \). Because \(K_{\mu\nu}^{(\xi)}\) is invertible
\begin{equation}\label{eq:defK}
	\left(K^{-1}\right)_{\mu\nu}^{(\xi)} = \frac{1}{k^2}\left( \delta_{\mu\nu} + (\xi - 1) \frac{k_\mu k_\nu}{k^2} \right),
\end{equation}
we can complete the square in (\ref{eq:quadraticForm})
\begin{equation}\label{eq:afterShift}
	S = \frac12 \int \frac{\dd[4]{k}}{(2\pi)^4} \bigg[ \hat{A}'_\mu(k)  K_{\mu\nu}^{(\xi)} \hat{A}'_\nu(-k) - \frac14 \hat{v}_\mu(k) \left(K^{-1}\right)_{\mu\nu}^{(\xi)} \hat{v}_\nu(-k) \bigg],
\end{equation}
with \(\hat{A}'_\mu(k) = \left( \hat{A}_\mu(k) - \frac12 \hat{v}_\rho(k) \left(K^{-1} \right)_{\mu\rho}^{(\xi)} \right)\). Since the path integral measure is invariant under the substitution above (i.e.\ \(\mathcal{D}\!\hat{A}' = \mathcal{D}\!\hat{A}\)), the functional integral (\ref{eq:pathInt}) now decouples: 
\begin{equation*}
	Z = \left( \int \mathcal{D}\!\hat{A}'\, e^{-S_{A'}} \right) \left( \int \mathcal{D}\hat{b}^-\, \mathcal{D}\hat{b}^+\, e^{-S_b} \right)= Z_{A'}Z_b ,
\end{equation*}
where \(S_{A'}\) is the first term of (\ref{eq:afterShift}) and \(S_b\) the second one. Since \(S_{A'}\) does not contain any dependence on the inter-plate distance \(L\), \(Z_{A'}\) will not give any contribution to the Casimir effect. We will thus treat it as an irrelevant (infinite) constant, and focus on \(Z_b\). 
Returning to the notation \(\hat{v}_\nu(k) = 2 i k_\mu \hat{H}_{\mu\nu}(k) \) in \(S_b\), we have 
\begin{equation*} 
	S_b = -\frac12 \int \frac{\dd[4]{k}}{(2\pi)^4} \bigg[ k_\mu \hat{H}_{\mu\rho}(k) \left(K^{-1}\right)_{\rho\sigma}^{(\xi)} k_\nu \hat{H}_{\nu\sigma}(-k) \bigg],
\end{equation*}
which we want to write as a quadratic form in the boundary fields. If we factor out the \(\hat{b}\)-field from \(\hat{H}_{\mu\nu}\), equation (\ref{eq:defHFourier}) becomes 
\begin{equation}\label{eq:defHH}
	\hat{H}_{\mu\nu}(k) = e^{i k_z z^\gamma} \underbrace{\Big[ (\delta_{i\mu}n_\nu - \delta_{i\nu}n_\mu)\cos(\theta^\gamma) +i \varepsilon_{i\mu\nu z}\sin(\theta^\gamma) \Big]}_{ H_{i\mu\nu}^\gamma} \hat{b}_i^\gamma(\mathbf{k}).
\end{equation}
Using the shorthand \(H_{i\mu\nu}^\gamma\), we can write the action compactly
\begin{align}\label{eq:4dSb}
	S_b &=  -\frac12 \int \frac{\dd[4]{k}}{(2\pi)^4} \bigg[ e^{i k_z (z^\gamma - z^\lambda)} H_{i\mu\rho}^\gamma H_{j\nu\sigma}^\lambda k_\mu k_\nu \left(K^{-1}\right)_{\rho\sigma}^{(\xi)} \hat{b}_i^\gamma(\mathbf{k}) \hat{b}_j^\lambda(-\mathbf{k}) \bigg] \nonumber \\
	&=-\frac12 \int \frac{\dd[3]{\mathbf{k}}}{(2\pi)^3} \bigg[ \hat{b}_i^\gamma(\mathbf{k}) \int \frac{\dd{k_z}}{2\pi} \Big[ e^{i k_z (z^\gamma - z^\lambda)} H_{i\mu\rho}^\gamma H_{j\nu\sigma}^\lambda k_\mu k_\nu \left(K^{-1}\right)_{\rho\sigma}^{(\xi)} \Big] \hat{b}_j^\lambda(-\mathbf{k}) \bigg].
\end{align}
As such, we have reached an explicitly Gaussian form for \(S_b\) as well; we only need to perform the \(k_z\)-integral. In order to do this, let us have a closer look at the tensor contractions in the \(k_z\)-integrand: 
\begin{align}
		H_{i\mu\rho}^\gamma H_{j\nu\sigma}^\lambda k_\mu k_\nu \left(K^{-1}\right)_{\rho\sigma}^{(\xi)} &= H_{i\mu\rho}^\gamma H_{j\nu\sigma}^\lambda k_\mu k_\nu \frac{\delta_{\rho\sigma}}{k^2} \label{eq:contraction1}\\ 
		&= \frac{1}{k^2} \Big( -\mathbf{k}^2 \mathbb{T}_{ij}(\mathbf{k}) \cos(\theta^\gamma - \theta^\lambda) +i \varepsilon_{ij\ell} k_\ell k_z \sin(\theta^\gamma - \theta^\lambda) \nonumber \\
		& \qquad\qquad\qquad\qquad\qquad\qquad\qquad\qquad\qquad +k^2\delta_{ij} \cos(\theta^\gamma) \cos(\theta^\lambda) \Big). \label{eq:contraction2}
\end{align}
In the first equality, we have made use of (\ref{eq:defK}) and the fact that \(H_{i\mu\nu}^\gamma\) is anti-symmetric in its last two indices. Notice that the gauge parameter \(\xi\) automatically disappears from our calculations, such that we have indeed found the promised gauge invariance of the effective 3D action. In the second equality, we have filled in the definition of \(H_{i\mu\nu}^\gamma\) and the angle difference formulas appear after a bit of algebra. \(\mathbb{T}_{ij}(\mathbf{k}) = \delta_{ij} - \frac{k_i k_j}{\mathbf{k}^2}\) denotes the 3-dimensional transversal projector, and one should note that the Levi-Civita symbol \(\varepsilon_{ij\ell}\) is a 3-dimensional one here. Let us already define the longitudinal projector \(\mathbb{L}_{ij}(\mathbf{k}) = \frac{k_i k_j}{\mathbf{k}^2}\) as well, since we will need it later on.

Inspecting (\ref{eq:contraction2}), we start by observing that the non duality-invariant last term does not contain any \(k_z\)-dependence, and as such gives rise to a factor \(\delta(z^\gamma-z^\lambda)\) when evaluating the \(k_z\)-integral in (\ref{eq:4dSb}). If \(\gamma \neq \lambda\), this factor simply equals zero, while for \(\gamma=\lambda\) we find \(\delta(0)\), which is also zero in dimensional regularization. As such, the boundary propagator will be duality-invariant, as expected. 

We can then perform the \(k_z\)-integral of the first two terms in (\ref{eq:contraction2}) using the integrals in Appendix \ref{ch:appendix}. Doing so, we arrive at the non-local 3D effective boundary action
\begin{equation}\label{eq:bndQuad0}
	S_b = \frac12 \int \frac{\dd[3]{\mathbf{k}}}{(2\pi)^3}  \hat{b}_i^\gamma(\mathbf{k}) \mathbb{K}_{ij}^{\gamma\lambda}(\mathbf{k}) \hat{b}_j^\lambda(-\mathbf{k}) ,
\end{equation}
with the quadratic form given by
\begin{equation}\label{eq:bndQuad}
	\mathbb{K}_{ij}^{\gamma\lambda}(\mathbf{k}) = \frac12 \Big( \abs{\mathbf{k}} \mathbb{T}_{ij} \cos(\theta^\gamma - \theta^\lambda) - \varepsilon^{\gamma\lambda}\varepsilon_{ij\ell} k_\ell \sin(\theta^\gamma - \theta^\lambda) \Big) e^{-\abs{\mathbf{k}}\abs{z^\gamma - z^\lambda}},
\end{equation} 
where we set \(\varepsilon^{-+} = 1\), and no summation over \(\gamma\) or \(\lambda\) is implied in the right-hand side. The effective action \eqref{eq:bndQuad0} clearly shows non-trivial boundary quantum dynamics, encoded in the ``classically trivial'' $b$-fields.

At this point, we notice that the \(b^\gamma\)-fields possess a gauge symmetry, which in momentum space is given by \(b_i^\gamma \rightarrow b_i^\gamma + \beta^\gamma(\mathbf{k}) k_i \) for two arbitrary functions \(\beta^\gamma\). We will fix this gauge by adding gauge fixing terms of the form \(b_i^-(\mathbf{k})\mathbb{L}_{ij}b_j^-(-\mathbf{k})\) and \(b_i^+(\mathbf{k})\mathbb{L}_{ij}b_j^+(-\mathbf{k})\) to the action, giving
\begin{equation}\label{eq:bndQuadGF}
	\mathbb{K}_{ij}^{\gamma\lambda}(\mathbf{k}) = \frac{\abs{\mathbf{k}}}{2} \bigg(  \eta^\gamma \delta^{\gamma\lambda}\mathbb{L}_{ij} +  \mathbb{T}_{ij} \cos(\theta^\gamma - \theta^\lambda) - \varepsilon^{\gamma\lambda}\varepsilon_{ij\ell} \frac{k_\ell}{\abs{\mathbf{k}}} \sin(\theta^\gamma - \theta^\lambda) \bigg) e^{-\abs{\mathbf{k}}\abs{z^\gamma - z^\lambda}},
\end{equation} 
where \(\eta^\gamma\) are the corresponding two gauge parameters, and again, no summation over \(\gamma\) or \(\lambda\) is implied in the right-hand side.

\section{The Casimir energy: two methods}\label{ch:methods}
\subsection{Casimir energy via the path integral}\label{ch:pathIntegral}
If we denote with \(\mathcal{E}\) the vacuum energy density, and with \(\mathcal{V}\) the (technically infinite) size of the spacetime integration volume, we can recall the fact that 
\begin{equation*}
	Z = e^{-\mathcal{E}\mathcal{V}}.
\end{equation*}
If we can find an expression for \(Z\), this gives us a way to calculate the Casimir energy of the system without the need to construct the energy-momentum tensor \(T_{\mu\nu}\). Since we are only interested in contributions to \(\mathcal{E}\) that depend on the inter-plate distance \(L\), we can simply use the Gaussian \(Z_b\):
\begin{equation*}
	Z_b = \frac{C}{\sqrt{\det(\mathbb{K})}},
\end{equation*}
for some irrelevant (infinite) constant \(C\). From this, we find for the vacuum energy density 
\begin{equation*}
	\mathcal{E} = \frac12 \ln(\det(\mathbb{K})),
\end{equation*}
where we have dropped the irrelevant infinite contributions of \(C\) and \(\mathcal{V}\).
If we write the functional determinant as the product of the (infinitely many) eigenvalues of \(\mathbb{K}\), we get
\begin{equation*}
	\mathcal{E} = \frac12 \int \frac{\dd[3]{\mathbf{k}}}{(2\pi)^3} \ln(\abs{\mathbb{K}(\mathbf{k})}),
\end{equation*}
where \(\abs{\mathbb{K}(\mathbf{k})}\) is the matrix determinant of the boundary propagator in momentum representation (\ref{eq:bndQuadGF}) (in contrast with the functional determinant \(\det(\mathbb{K})\)). We find that\footnote{The easiest way to calculate this determinant is by writing \(\mathbb{K}\) as a sum of three rank-1 projectors with orthogonal eigenvectors: see equation (\ref{eq:bndQuadProj}) below.} 
\begin{equation*}
	\abs{\mathbb{K}(\mathbf{k})} = \left( \frac{\abs{\mathbf{k}}}{2} \right)^6 \eta^- \eta^+ \left( 1- 2 \cos(2\delta) e^{-2\abs{\mathbf{k}}L} + e^{-4\abs{\mathbf{k}}L} \right),
\end{equation*}
where again \(\delta = \theta^+ - \theta^-\).
Only the last factor is \(L\)-dependent, so we are left with
\begin{align*}
	\mathcal{E} &= \frac12 \int \frac{\dd[3]{\mathbf{k}}}{(2\pi)^3} \ln\left( 1- 2 \cos(2\delta) e^{-2\abs{\mathbf{k}}L} + e^{-4\abs{\mathbf{k}}L} \right) \\
	&= \frac{1}{4\pi^2} \int_0^{\infty} \dd{k} k^2 \ln\left( 1- 2 \cos(2\delta) e^{-2kL} + e^{-4kL} \right),
\end{align*}
where we have switched to spherical coordinates and integrated out the angular part. The radial integral can be computed analytically to yield
\begin{equation}\label{eq:CasEPathIntegral}
	\mathcal{E} = -\frac{\Re \textrm{Li}_4\left(e^{2i\delta}\right)}{8\pi^2 L^3},
\end{equation}
where \(\textrm{Li}_4\) is the polylogarithm of order 4. The Casimir force per unit area is then given by
\begin{equation*}
	F = -\frac{\partial \mathcal{E}}{\partial L} = -\frac{3\Re \textrm{Li}_4(e^{2i\delta})}{8\pi^2 L^4}.
\end{equation*}
This value was first found in \cite{Rode:2017yqy} using reflection matrices to construct the Green's tensor. Notice that the Casimir force can be attractive, repulsive or zero depending on the value of \(\delta\). A more complete discussion of its behavior can be found in \cite{Rode:2017yqy,Canfora:2022xcx}.

\subsection{Casimir energy via the energy-momentum tensor}\label{ch:EMT}
In this section we explore a second method for calculating the Casimir force, using the energy-momentum tensor (EMT) \cite{Bordag:1983zk,Bordag:2009zz}. This method is better suited for studying moving boundaries and the dynamical Casimir effect \cite{Bordag:1985rb}. Another merit is that it shows that the EMT has some contributions from the boundary in the general PEMC case, something not noted before to the best of our knowledge\footnote{Although the potential implications of surface contributions in the tensor's renormalization are known, see e.g.~\cite{Romeo:2000wt,Kennedy:1979ar,Deutsch:1978sc}.}. Indeed, one could be inclined to simply use the improved Maxwell EMT 
\begin{equation}\label{eq:MaxwellEMT}
	T_{\mu\nu}^{\text{Maxwell}} = F_{\mu\rho}F_{\nu\rho} - \frac14 \delta_{\mu\nu} F_{\rho\sigma} F_{\rho\sigma}.
\end{equation}
Doing so, one in fact ignores the presence of the boundaries. It is necessary, however, to be more cautious and to take full account of the boundary. Indeed, we verified that working with \eqref{eq:MaxwellEMT} would lead to a wrong expression for the Casimir energy in the general case.

One possibility is to construct the Bessel-Hagen EMT \cite{Baker:2021hly,Freese_2022} for our Lagrangian density (\ref{eq:action}) (without the gauge fixing term). Alternatively, one could use the canonical Noether EMT \cite{Noether1918} followed by the Belinfante improvement procedure \cite{Belinfante1940OnTC}. Both give the same tensor 
\begin{equation}\label{eq:EMT}
	T_{\mu\nu} = (F_{\mu\rho} + H_{\mu\rho})F_{\nu\rho} - \delta_{\mu\nu} \left( \frac14 F_{\rho\sigma} + \frac12 H_{\rho\sigma} \right) F_{\rho\sigma}.
\end{equation}
Another option is to construct the Hilbert EMT by varying the action with respect to the metric tensor (e.g.\ \cite{Padmanabhan:2010zzb}), which yields (\ref{eq:EMT}) but symmetrized w.r.t.\ \(\mu\) and \(\nu\) (as one expects from an EMT). Since we will only use diagonal elements however, we will simply continue with (\ref{eq:EMT}). Note that we indeed have picked up boundary contributions compared to the Maxwell EMT (\ref{eq:MaxwellEMT}). These contributions modify the Casimir energy. More specifically, they ensure that the result is duality-invariant (as it should), as such making manifest their physical necessity.

We can now write for the vacuum energy density (the minus sign is due to the Wick rotation to Euclidean space)
\begin{equation*}
	\mathcal{E} = -\int \dd{z} \left\langle T_{00}(x)\right\rangle.
\end{equation*}
We will again first integrate out the photon field by performing the shift \(A \rightarrow A'\) (cfr.\ (\ref{eq:afterShift})). Looking at the form (\ref{eq:EMT}) of \(T_{\mu\nu}\), we would like to know how the field strength \(F_{\mu\nu}\) transforms under this shift. We find that 
\begin{equation}\label{eq:shiftF}
	F_{\mu\nu} \rightarrow F_{\mu\nu} + G_{\mu\nu},
\end{equation}
where in Fourier space \(G_{\mu\nu}\) is given by
\begin{equation}\label{eq:defG}
	\hat{G}_{\mu\nu}(k) = \frac{k_\rho}{k^2}\left( k_\mu \hat{H}_{\nu\rho}(k) - k_\nu \hat{H}_{\mu\rho}(k) \right),
\end{equation}
with \(\hat{H}_{\mu\nu}\) given by (\ref{eq:defHH}). Using (\ref{eq:shiftF}), we can now easily write down the energy density after the shift
\begin{align*}
	\mathcal{E} &= -\int \dd{z} \left\langle \big((F_{\mu\rho} + G_{\mu\rho}) + H_{\mu\rho}\big)(F_{\nu\rho}+G_{\nu\rho}) - \delta_{\mu\nu} \left(  \frac14 (F_{\rho\sigma} + G_{\rho\sigma}) + \frac12 H_{\rho\sigma}\right)(F_{\rho\sigma} + G_{\rho\sigma}) \right\rangle \Bigg|_{\mu=\nu=0} \\
	&= -\int \dd{z} \left\langle T_{\mu\nu}^{\text{Maxwell}} + \underbrace{(G_{\mu\rho} + H_{\mu\rho})G_{\nu\rho} - \delta_{\mu\nu} \left( \frac14 G_{\rho\sigma} + \frac12 H_{\rho\sigma} \right) G_{\rho\sigma}}_{T_{\mu\nu}^\text{bnd}} \right\rangle\Bigg|_{\mu=\nu=0},
\end{align*}
where in the second equality we have used that only the terms quadratic in the \(A\)-fields (resp.\ \(b\)-fields) have non-zero expectation value after the shift. The first term only contains the now-decoupled photon field, so it does not depend on the inter-plate distance \(L\), meaning we can safely drop it (this corresponds to dropping \(Z_{A'}\) from \(Z\) in section \ref{ch:pathIntegral}).  

The straightforward way to proceed would be to set \(\mu=\nu=0\), but when calculating \(\langle G_{0\rho} G_{0\rho} \rangle\) and \(\langle H_{0\rho}G_{0\rho} \rangle\), we found the uncontracted 0-indices to be bothersome. To circumvent this problem, introduce the tensor \(\Theta_{\mu\nu}\), defined by
\begin{equation*}
	\int \dd{z} T^\text{bnd}_{\mu\nu} = b_i^\gamma \left(\Theta_{\mu\nu}\right)_{ij}^{\gamma\lambda} b_j^\lambda.
\end{equation*}
After the photon field has been integrated out, the only objects left that carry spacetime indices are the Euclidean metric \(\delta_{\mu\nu}\) and the normal vector \(n_\mu\). As such, \(\Theta_{\mu\nu}\) can only consist of two components
\begin{equation*}
	\Theta_{\mu\nu} = \Upsilon \delta_{\mu\nu} + \Xi n_\mu n_\nu.
\end{equation*}
Recalling that Latin indices \(\in \{t,x,y\}\) and that $n_\mu=\delta_{\mu z}$, we can then use the trick that
\begin{equation*}
	\mathcal{E} = -\int \dd{z} \left\langle T^\text{bnd}_{00}\right \rangle = -\left(\Theta_{00}\right)_{jk}^{\gamma\lambda} \left\langle b_j^\gamma b_k^\lambda \right\rangle= -\Upsilon_{jk}^{\gamma\lambda} \left\langle b_j^\gamma b_k^\lambda \right\rangle = -\frac13 \left(\Theta_{ii}\right)_{jk}^{\gamma\lambda} \left\langle b_j^\gamma b_k^\lambda \right\rangle = -\frac13 \int \dd{z} \left\langle T^\text{bnd}_{ii} \right\rangle,
\end{equation*}
which gets rid of the uncontracted 0-indices. 
We thus find
\begin{equation}\label{eq:Etrick13}
	\mathcal{E} = -\int \dd{z} \left\langle \frac13 (G_{i\rho} + H_{i\rho})G_{i\rho} - \left( \frac14 G_{\rho\sigma} + \frac12 H_{\rho\sigma} \right) G_{\rho\sigma} \right\rangle.
\end{equation}
In order to continue, we need an expression for the \(b\)-propagators. In Fourier space we have
\begin{equation}\label{eq:prop}
	\left\langle \hat{b}_i^\gamma(\mathbf{k}) \hat{b}_j^\lambda(\mathbf{q}) \right\rangle = (2\pi)^3 \delta^{(3)}(\mathbf{k}+\mathbf{q}) \left( \mathbb{K}^{-1} \right)_{ij}^{\gamma\lambda}(\mathbf{k}).
\end{equation}
Let us now consider the vacuum energy density (\ref{eq:Etrick13}). We will start with the third term. Going to Fourier space and filling in (\ref{eq:defG}) for \(\hat{G}\), one recognizes \(\mathbb{K}\) in the form of the \(k_z\)-integral in (\ref{eq:4dSb}), contracted with the propagator \(\langle b_i^\gamma(\mathbf{k})b_j^\lambda(\mathbf{q}) \rangle\). Writing this out yields
\begin{equation*}
	-\frac14 \int \dd{z} \langle G_{\rho\sigma}(x)G_{\rho\sigma}(x) \rangle =  \frac12 \int \frac{\dd[3]{\mathbf{k}}}{(2\pi)^3} \mathbb{K}_{ji}^{\lambda\gamma} \left( \mathbb{K}^{-1} \right)_{ij}^{\gamma\lambda} = 3 \int \frac{\dd[3]{\mathbf{k}}}{(2\pi)^3},
\end{equation*}
which is an infinite constant independent of the inter-plate distance \(L\), and thus irrelevant for our purpose. Analogously, the last term \(-\frac12 \int \dd{z} \langle H_{\rho\sigma}(x)G_{\rho\sigma}(x) \rangle\) can be seen to be irrelevant as well. We are thus left with
\begin{equation*}
	\mathcal{E} = -\frac13 \int \dd{z} \left\langle G_{i\rho}G_{i\rho} + H_{i\rho}G_{i\rho} \right\rangle.
\end{equation*}
Let us calculate both terms in Fourier space. We find
\begin{align*}
	-\frac13 \int \dd{z} \left\langle G_{i\rho}G_{i\rho} \right\rangle = -\frac13 \int \frac{\dd[4]{k}}{(2\pi)^4} \Bigg[&\frac{\mathbf{k}^2}{k^4} \langle k_\mu \hat{H}_{\rho\mu}(k) k_\nu \hat{H}_{\rho\nu}(-k) \rangle \\  
	+ &\frac{1}{k^2} \langle k_\mu \hat{H}_{i\mu}(k) k_\nu \hat{H}_{i\nu}(-k) \rangle \Bigg]
\end{align*}
and
\begin{align*}
	-\frac13 \int \dd{z} \left\langle H_{i\rho}G_{i\rho} \right\rangle = \frac13 \int \frac{\dd[4]{k}}{(2\pi)^4} \Bigg[&\frac{1}{k^2} \langle k_\mu \hat{H}_{\rho\mu}(k) k_i \hat{H}_{\rho i}(-k) \rangle \\
	+ &\frac{1}{k^2} \langle k_\mu \hat{H}_{i \mu}(k) k_\nu \hat{H}_{i \nu}(-k) \rangle \Bigg].
\end{align*}
The last terms of both expressions are identical up to a sign, and thus neatly cancel each other. (Also note that they are divergent, such that we are regularizing.) Let us then calculate the VEV of the left-over terms. To do this, factor out the \(\hat{b}\)-field from each \(\hat{H}_{\mu\nu}\) factor using (\ref{eq:defHH}), then fill in the propagator (\ref{eq:prop}), and perform contractions similar to the one in (\ref{eq:contraction1}). For the first term we find
\begin{align}\label{eq:term1}
	\begin{split}
		& -\frac13 \int \frac{\dd[4]{k}}{(2\pi)^4} \frac{\mathbf{k}^2}{k^4} \langle k_\mu \hat{H}_{\rho\mu}(k) k_\nu \hat{H}_{\rho\nu}(-k) \rangle \\ &= -\frac13 \int \frac{\dd[4]{k}}{(2\pi)^4} \frac{\mathbf{k}^2}{k^4} \left(\mathbb{K}^{-1}\right)_{ij}^{\gamma\lambda}(-\mathbf{k}) \Bigg[ -\mathbf{k}^2 \mathbb{T}_{ij} \cos(\theta^\gamma - \theta^\lambda)  \\
		&\hspace{50pt} + i\varepsilon_{ij\ell} k_\ell k_z \sin(\theta^\gamma - \theta^\lambda) + k^2 \delta_{ij} \cos(\theta^\gamma)\cos(\theta^\lambda)\Bigg] e^{i k_z (z^\gamma - z^\lambda)},
	\end{split}
\end{align}
where the last term is not duality-invariant. Recall that in the calculation of the effective action (\ref{eq:contraction2}), we could set this term to zero in our regularization scheme. Here, it does not disappear however. Also note that the \(k_z\)-integral is finite here, and needs no further regularization.
For the second term we find in a similar manner
\begin{align}\label{eq:term2}
	\begin{split}
		&\frac13 \int \frac{\dd[4]{k}}{(2\pi)^4} \frac{1}{k^2} \langle k_\mu \hat{H}_{\rho\mu}(k) k_i \hat{H}_{\rho i}(-k) \rangle \\ &= -\frac13 \int \frac{\dd[4]{k}}{(2\pi)^4} \frac{1}{k^2} \left(\mathbb{K}^{-1}\right)_{ij}^{\gamma\lambda}(-\mathbf{k}) \Bigg[ \mathbf{k}^2 \mathbb{T}_{ij} \cos(\theta^\gamma - \theta^\lambda) \\
		&\hspace{50pt} - \frac{i}{2} \varepsilon_{ij\ell} k_\ell k_z \sin(\theta^\gamma - \theta^\lambda) - \mathbf{k}^2 \delta_{ij} \cos(\theta^\gamma)\cos(\theta^\lambda)\Bigg] e^{i k_z (z^\gamma - z^\lambda)},
	\end{split}
\end{align}
where the same duality breaking term appears, but with an opposite sign. As such, we retrieve a duality-invariant expression for \(\mathcal{E}\), as expected. Adding (\ref{eq:term1}) and (\ref{eq:term2}) together, we get
\begin{align}
	\mathcal{E} &= -\frac13 \int \frac{\dd[4]{k}}{(2\pi)^4} \left(\mathbb{K}^{-1}\right)_{ij}^{\gamma\lambda}(-\mathbf{k}) \Bigg[ \left( -\frac{\mathbf{k}^4}{k^4} + \frac{\mathbf{k}^2}{k^2} \right) \mathbb{T}_{ij} \cos(\theta^\gamma - \theta^\lambda) \nonumber\\
	&\hspace{6cm} +i\left( \frac{k_z \mathbf{k}^2}{k^4} -\frac{k_z}{2k^2} \right) \varepsilon_{ij\ell} k_\ell \sin(\theta^\gamma - \theta^\lambda) \Bigg] e^{i k_z (z^\gamma - z^\lambda)} \nonumber \\
	&= -\frac16 \int \frac{\dd[3]{\mathbf{k}}}{(2\pi)^3} \left(\mathbb{K}^{-1}\right)_{ij}^{\gamma\lambda}(-\mathbf{k}) \left[ \mathbb{K}_{ij}^{\gamma\lambda}(\mathbf{k}) - \abs{\mathbf{k}} \abs{z^\gamma - z^\lambda} \mathbb{K}_{ij}^{\gamma\lambda}(\mathbf{k}) \right], \label{eq:k3Int}
\end{align}
where in the second line, after having performed the \(k_z\)-integral using Appendix \ref{ch:appendix}, we recognize \(\mathbb{K}_{ij}^{\gamma\lambda}\) in the form of (\ref{eq:bndQuad}). 

In order to finish, we still need an expression for \(\mathbb{K}^{-1}\). The easiest way to find this inverse is by introducing rank-1 projectors
\begin{align*}
	P^1_{ij} &= \frac12 \left( \mathbb{T}_{ij} + i \varepsilon_{ij\ell} \frac{k_\ell}{\abs{\mathbf{k}}} \right),\\
	P^2_{ij} &= \frac12 \left( \mathbb{T}_{ij} - i \varepsilon_{ij\ell} \frac{k_\ell}{\abs{\mathbf{k}}} \right)
\end{align*}
such that the gauge-fixed quadratic operator (\ref{eq:bndQuadGF}) becomes 
\begin{equation}\label{eq:bndQuadProj}
	\mathbb{K}_{ij}^{\gamma\lambda}(\mathbf{k}) = \frac{\abs{\mathbf{k}}}{2} \bigg(  \eta^\gamma \delta^{\gamma\lambda}\mathbb{L}_{ij} +  P^1_{ij} e^{i\varepsilon^{\gamma\lambda} (\theta^\gamma - \theta^\lambda)} + P^2_{ij} e^{-i\varepsilon^{\gamma\lambda} (\theta^\gamma - \theta^\lambda)} \bigg) e^{-\abs{\mathbf{k}}\abs{z^\gamma - z^\lambda}}.
\end{equation} 
We can then easily write down the inverse operator
\begin{equation}\label{eq:inverseK}
	\left( \mathbb{K}^{-1}\right)_{ij}^{\gamma\lambda}(\mathbf{k}) = \frac{2}{\abs{\mathbf{k}}} \bigg( \frac{1}{\eta^\gamma} \delta^{\gamma\lambda} \mathbb{L}_{ij} + P^1_{ij} A^{\gamma\lambda} + P^2_{ij} \bar{A}^{\gamma\lambda} \bigg),
\end{equation}
where 
\begin{align*}
	A^{\gamma\lambda} &= \frac{1}{e^{-2\abs{\mathbf{k}}L - 2i \delta}-1} 
		\begin{pmatrix}
			-1 & e^{-\abs{\mathbf{k}}L - i\delta} \\
			e^{-\abs{\mathbf{k}}L - i\delta} & -1 
		\end{pmatrix} \\
	\bar{A}^{\gamma\lambda} &= 	A^{\gamma\lambda}(\delta \rightarrow -\delta)
\end{align*}
with \(\delta = \theta^+ - \theta^-\) and \(L = z^+ - z^-\).

Now we can evaluate the integral (\ref{eq:k3Int}). Let us start with the first term. Inspecting expression (\ref{eq:inverseK}) for \(\mathbb{K}^{-1}\), we see that it is symmetric in the plate-indices, and that transposing the spacetime indices is equivalent to sending \(\mathbf{k} \rightarrow - \mathbf{k}\), resulting in \(\left( \mathbb{K}^{-1}\right)_{ij}^{\gamma\lambda}(-\mathbf{k}) = \left( \mathbb{K}^{-1}\right)_{ji}^{\lambda\gamma}(\mathbf{k})\). This means the first term in (\ref{eq:k3Int}) is proportional to \(\left( \mathbb{K}^{-1}\right)_{ji}^{\lambda\gamma}(\mathbf{k}) \mathbb{K}_{ij}^{\gamma\lambda}(\mathbf{k}) = 6\), which is independent of \(L\) and thus irrelevant. For the second term, the factor \(\abs{z^\gamma - z^\lambda}\) is zero if \(\gamma = \lambda\), and \(L\) otherwise, such that we get
\begin{align*}
	\mathcal{E} &= \frac{L}{3} \int \frac{\dd[3]{\mathbf{k}}}{(2\pi)^3} \abs{\mathbf{k}} \left(\mathbb{K}^{-1}\right)_{ji}^{+-}(\mathbf{k})  \mathbb{K}_{ij}^{-+}(\mathbf{k})  \\
	&= \frac{L}{3} \int \frac{\dd[3]{\mathbf{k}}}{(2\pi)^3}  \abs{\mathbf{k}}  \left( e^{-i\delta} A^{+-} + e^{i\delta}\bar{A}^{+-} \right)e^{-\abs{\mathbf{k}} L} \\
	&= \frac{L}{3} \int \frac{\dd[3]{\mathbf{k}}}{(2\pi)^3}  \abs{\mathbf{k}} \left( \frac{1}{1-e^{2\abs{\mathbf{k}}L+2i\delta}} +  \frac{1}{1-e^{2\abs{\mathbf{k}}L-2i\delta}} \right) \\
	&= -\frac{\Re\textrm{Li}_4\left( e^{2i\delta} \right)}{8\pi^2 L^3},
\end{align*}
which indeed equals the result (\ref{eq:CasEPathIntegral}) from section \ref{ch:pathIntegral}.

Let us finish with a comment for the specific case in which both plates are PECs (i.e.\ \(\theta^+=\theta^-=\frac{\pi}{2}\)). In this case, starting from the ordinary Maxwell EMT (\ref{eq:MaxwellEMT}) still gives the correct result \(\mathcal{E} = -\frac{\pi^2}{720L^3}\) (as was done in \cite{Bordag:1983zk}). As such, one only seems to encounter the need for boundary contributions to the EMT when studying the more general PEMC boundary conditions. Indeed, in the PEMC case, the Casimir expression picks up an incorrect duality-breaking term \(\propto \cos(\theta^\gamma)\cos(\theta^\lambda)\) when using the Maxwell EMT (\ref{eq:MaxwellEMT}).

\section{Conclusion and outlook}\label{ch:conclusion}
We rederived the Casimir energy for two parallel, infinite, infinitely thin, perfect electromagnetic conductors. As a first step, we implemented these boundary conditions in an explicitly gauge-invariant manner by introducing Lagrange multiplier fields into the action. Subsequently, the invariance of the full action under electromagnetic duality transformations was shown. After this, a 3D non-local effective boundary theory was obtained by integrating out the photon field. We then discussed two related functional integral methods to calculate the Casimir energy. In the first method, it was derived directly from the path integral of the boundary theory, resulting in the usual log-det formula. In the second method, we calculated the Casimir energy as the vacuum expectation value of the \(00\)-component of the energy-momentum tensor. Remarkably, new boundary contributions to the EMT needed to be taken into account for this. 

Identifying the correct energy-momentum tensor is of the utmost importance for cases in which the local energy density is necessary to compute the Casimir energy upon integration over the relevant geometry, e.g.\ with dynamical boundary conditions. We hope to come back to this in future work. 

Clearly, the auxiliary boundary fields formalism offers several benefits: not only do these fields allow to deal with the boundary conditions in an explicitly gauge invariant manner, it also makes quite clear that field dependent quantities, such as the EMT, can and do receive boundary corrections, which we explicitly identified.

Another extension of the current work will be to add the nonlinear boundary conditions of Yang-Mills theories, a topic that recently attracted attention in e.g.\ \cite{Chernodub:2023dok}. We already noted that the boundary fields become dynamically non-trivial when integrating out the bulk modes and we foresee a potentially interesting connection with the ``glueton boundary mode'' introduced in the lattice QCD work \cite{Chernodub:2023dok}. 

\section*{Acknowledgments}
The work of D.D.~and T.O.~was supported by KU Leuven IF project C14/21/087. The work of S.S.~was funded by FWO PhD-fellowship fundamental research (file number: 1132823N).

\appendix
\section{Fourier conventions and useful integrals}\label{ch:appendix}
We will follow the Fourier convention used in \cite{Peskin:1995ev}
\begin{equation*}
	X_i(x) = \int \frac{\dd[d]{k}}{(2\pi)^d} \hat{X}_i(k)e^{-ik\cdot x}
\end{equation*}
for a \(d\)-dimensional vector field \(X\). In this convention, the Dirac delta in \(H_{\mu\nu}\) (\ref{eq:defH}) is Fourier transformed as
\begin{equation*}
	\delta(z-z^\gamma) = \int \frac{\dd{k_z}}{2\pi} e^{-ik_z(z-z^\gamma)}.
\end{equation*}

Let us list the \(k_z\)-integrals we encounter in our calculations
\begin{align*}
	\begin{split}
		&\int \frac{\dd{k_z}}{2\pi} \frac{e^{i k_z(z^\gamma - z^\lambda)} }{k^2} = \frac{1}{2\abs{\mathbf{k}}}e^{-\abs{\mathbf{k}} \abs{z^\gamma - z^\lambda}}, \\
		&\int \frac{\dd{k_z}}{2\pi} \frac{k_z e^{i k_z(z^\gamma - z^\lambda)} }{k^2} = 
		-\frac{i}{2} \varepsilon^{\gamma\lambda} e^{-\abs{\mathbf{k}} \abs{z^\gamma - z^\lambda}}, \\
		&\int \frac{\dd{k_z}}{2\pi} \frac{e^{i k_z(z^\gamma - z^\lambda)} }{k^4} = \frac{1+\abs{\mathbf{k}} \abs{z^\gamma - z^\lambda}}{4\abs{\mathbf{k}}^3}  e^{-\abs{\mathbf{k}} \abs{z^\gamma - z^\lambda}}, \\
		&\int \frac{\dd{k_z}}{2\pi} \frac{k_z e^{i k_z(z^\gamma - z^\lambda)} }{k^4} = 
		-\frac{i}{4} \varepsilon^{\gamma\lambda} \frac{\abs{z^\gamma - z^\lambda}}{\abs{\mathbf{k}}} e^{-\abs{\mathbf{k}} \abs{z^\gamma - z^\lambda}},
	\end{split}
\end{align*} 
where we set \(\varepsilon^{-+} = 1\), and no summation over \(\gamma\) or \(\lambda\) is implied in the right-hand side. 

\bibliography{bibliography}

\end{document}